\documentclass[prd,aps,superscriptaddress,twocolumn,nofootinbib,floatfix,longbibliography,10pt]{revtex4-2}
\usepackage{graphicx}% Include figure files
\usepackage{dcolumn}% Align table columns on decimal point
\usepackage{bm}% bold math
\usepackage{subcaption}
\usepackage{float}
\usepackage{multirow}
\usepackage{booktabs}
\usepackage{amsmath}
\usepackage{cases}
\usepackage{latexsym,amssymb}
\usepackage[mathscr]{eucal}
\usepackage[switch*]{lineno}
\usepackage{array}
\usepackage{cuted}
\usepackage{comment}
\usepackage[bookmarks=true,
   colorlinks=true,
   linkcolor=blue,
   urlcolor=blue,
   citecolor=blue,
   bookmarks=true,
   hyperindex=true
]{hyperref}

\begin{document}

\title{Matter-induced global regularity in non-polynomial quasi-topological gravity with Born-Infeld electrodynamics}%{Globally regular charged black holes in non-polynomial quasi-topological gravity with Born-Infeld electrodynamics}

\author{Hong-Lin Liu}
\email{hlliuphy@163.com}
\affiliation{School of Physics and Astronomy, China West Normal University, Nanchong 637009, China}

\author{Zhong-Wen Feng}
\email{zwfengphy@cwnu.edu.cn}
\affiliation{School of Physics and Astronomy, China West Normal University, Nanchong 637009, China}

\author{Qing-Quan Jiang}
\email{qqjiangphys@yeah.net}
\affiliation{School of Physics and Astronomy, China West Normal University, Nanchong 637009, China}

\author{Xia Zhou}
\email{zhoux@my.swjtu.edu.cn}
\affiliation{School of Physics and Astronomy, China West Normal University, Nanchong 637009, China}

\author{Xue-Ling Mu}
\email{muxueling@cdu.edu.cn}
\affiliation{College of Electronic Information and Electrical Engineering, Chengdu University, Chengdu, 610106, China}


\begin{abstract}
We construct exact static, spherically symmetric charged solutions in four-dimensional non-polynomial quasi-topological gravity coupled to Born-Infeld electrodynamics. We focus on the model $h\left(p\right)=p/(1+\ell^{2}p)$, whose vacuum branch develops a curvature singularity at a finite radius. We show that Born-Infeld nonlinearities can remove this singularity within a finite region of parameter space, yielding globally regular geometries with an asymptotically flat exterior and a finite-curvature AdS-type core. The regular sector contains both horizonless configurations and regular black holes, separated by a degenerate-horizon boundary. We further identify a continuous regular black hole branch with a triple-degenerate inner horizon and a simple outer event horizon, satisfying $\kappa_-=0$ and $\kappa_+\neq0$. These results provide a converse example to cases in which introducing charge spoils the regularity of a black hole that is regular in vacuum.
\end{abstract}

\maketitle
\allowdisplaybreaks[4]
% \linenumbers

\section{Introduction}
\label{sec1}

Black holes are among the most remarkable predictions of general relativity and provide a natural arena for probing gravity in the strong-field regime. However, classical black hole solutions inevitably exhibit spacetime singularities in their interiors, which are generally regarded as signaling the breakdown of the classical description in regions of extremely high curvature~\cite{Penrose:1964wq,Hawking:1970zqf}. Regular black holes (RBHs) have therefore been proposed as nonsingular alternatives to classical black holes, replacing the singular interior with a regular region of finite curvature while preserving the standard black hole geometry at large distances~\cite{Ansoldi:2008RegularReview,Frolov:2016Regular,Lan:2023Review,CarballoRubio:2025Paradigm,Bueno:2025zaj,Borissova:2026Regular4D}. Since the pioneering proposal by Bardeen, a variety of RBHs  have been extensively investigated~\cite{Bardeen:1968,Dymnikova:1992ux,Borde:1996Topology,Hayward:2005gi,Feng:2024BlackWhite,Feng:2025SymmetricBlackWhite,Diez:2026bmp}. The existence of such spacetimes does not by itself explain how they arise from an underlying theory. A more fundamental question is whether RBHs can emerge as natural solutions of a well-defined theory of gravity rather than being introduced through phenomenological modifications of the metric.

One important route toward obtaining RBHs from an underlying theory is to introduce matter sources with suitable strong-field properties, among which nonlinear electrodynamics (NED) has been extensively studied. Its nonlinear behavior at short distances has led to a variety of charged RBHs in Einstein gravity~\cite{AyonBeato:1998ub,AyonBeato:1999rg,AyonBeato:2000zs,Bronnikov:2000vy,Burinskii:2002pz,Dymnikova:2004zc,AyonBeato:2004ih,Balart:2014cga,Fan:2016hvf,Li:2023AnalyticNED,Bronnikov:2022NEDReview}. Such constructions depend sensitively on the choice of NED model and are subject to constraints from the weak-field limit, energy conditions, and central regularity~\cite{Balart:2014jia,Bronnikov:2017sgg,Toshmatov:2018Comment,Bokulic:2022Constraints,Bokulic:2024ReverseEngineering,Bokulic:2026Conundrum,Wang:2026jvo,Wang:2026sqr}. Regularization through nonlinear electromagnetic fields alone is therefore not a generic mechanism, which also motivates the study of modified gravitational dynamics in the high-curvature regime. Quasi-topological gravity (QTG) and generalized quasi-topological gravity (GQTG) provide a useful class of higher-curvature theories for this purpose. Their static, spherically symmetric field equations retain a simple structure, while such theories can be constructed at arbitrary orders in curvature~\cite{Colleaux:2017ibe,OlivaRay:2010QTG,Myers:2010QTG,Bueno:2016ECG,Bueno:2016FourDBH,Hennigar:2017GQTG,Bueno:2017HigherDerivativeBH,Bueno:2019HigherCurvature,Bueno:2020AllOrders,Bueno:2023WholeShebang,Bueno:2026BirkhoffQTG}. In theories containing an infinite tower of higher-curvature corrections, RBHs have been obtained as vacuum solutions of purely gravitational theories~\cite{Bueno:2025RegularPureGravity,Frolov:2025RegularQTG,Bueno:2025ThinShellCollapse,Bueno:2025DynamicalFormation,Tsuda:2026FanWang,Ling:2026BigBounce,Mazza:2026taj}. This construction has also been realized in four-dimensional non-polynomial quasi-topological gravity (NPQTG), where single-function static, spherically symmetric solutions are determined by an algebraic equation and vacuum RBHs can arise from explicit four-dimensional covariant gravitational actions~\cite{Bueno:2025zaj,Borissova:2026Regular4D}. A more general formulation based on an integrable two-dimensional Horndeski reduction has further extended the class of spherical QTG theories~\cite{Borissova:2026krh}.

It is worth noting that regular vacuum black hole solutions do not necessarily remain regular once charge or matter fields are introduced. Charged solutions can obey more restrictive regularity conditions than their uncharged counterparts, and in some cases charge leads to finite-radius curvature singularities~\cite{CarballoRubio:2026Charging,Cano:2020ElectromagneticQTG,Cano:2020RNSingularities}. Similar behavior occurs in QTG coupled to matter, where the global properties of the matter-coupled solution cannot in general be inferred from the corresponding vacuum geometry~\cite{Hennigar:2025yqm,Hao:2025utc,Bueno:2026MatterQTG}. In particular, recent work on QTG coupled to Born-Infeld NED found that some models with regular vacuum black holes become singular after charge is introduced, whereas others remain regular~\cite{PinedoSoto:2026QTGBI}. Previous studies have therefore mainly addressed whether regular vacuum black holes remain regular after charge or matter fields are added~\cite{CarballoRubio:2026Charging,Bueno:2026MatterQTG,PinedoSoto:2026QTGBI,Colleaux:2026hat}. Here we consider the complementary situation in which the vacuum solution of a higher-curvature theory is singular at a finite radius. \textit{We ask whether nonlinear matter can alter the radial dependence of the gravitational field equation so that the charged geometry avoids this singularity and becomes globally regular.}

To study this problem, we consider the single-function static, spherically symmetric sector of four-dimensional NPQTG~\cite{Borissova:2026Regular4D}. Introducing the curvature variable $p=[1-f\left(r\right)]/r^{2}$, the Hayward regular vacuum geometry corresponds, in the normalization adopted here, to $h_{\rm H}\left(p\right)=p/(1-\ell^{2}p)$~\cite{Borissova:2026Regular4D,PinedoSoto:2026QTGBI}. We instead consider $h\left(p\right)={p}/({1+\ell^{2}p})$. The two choices have the same Einstein limit at low curvature but correspond to different higher-curvature corrections and therefore define distinct NPQTG models. Unlike the Hayward-type choice, the present model remains singular in vacuum. As the radius decreases, the vacuum field equation reaches the critical value $1/\ell^{2}$ at a finite radius, where the characteristic relation becomes singular and the curvature diverges. We then couple the theory to Born-Infeld nonlinear electrodynamics, which recovers Maxwell theory in the weak-field limit while modifying the electromagnetic dynamics in the strong-field regime~\cite{Born:1934Infeld,Gibbons:2001BornInfeld,Sorokin:2022NED}. This provides a setting in which to determine whether Born-Infeld nonlinearities can prevent the charged solution from reaching the critical value and restore global regularity. The resulting geometries can also be examined through the geometric convergence conditions entering the singularity theorems, which in higher-curvature gravity should be evaluated directly on shell rather than inferred from matter energy conditions~\cite{Borissova:2026prd}.

Even if the charged solution becomes globally regular, two further questions remain. The first is whether regularization persists over a finite region of parameter space and whether the regular solutions possess black hole horizons or remain horizonless. The second concerns the inner Cauchy horizon of RBHs. For a nondegenerate inner horizon, the infinite blueshift is closely related to the standard mass-inflation instability~\cite{PoissonIsrael:1989,PoissonIsrael:1990,Ori:1991,Dafermos:2003CauchyHorizon,CarballoRubio:2018Viability,CarballoRubio:2021InnerHorizon,DiFilippo:2022InnerHorizon}. Previous studies have suggested that, when the inner-horizon surface gravity vanishes, $\kappa_-=0$, while the outer horizon remains nondegenerate, the standard exponential blueshift near the inner horizon may be altered~\cite{CarballoRubio:2022InnerExtremal,DiFilippo:2025InnerExtremal,Feng:2026SubPlanckian,Gao:2025plm,Borissova:2026prd}. This condition does not establish the absence of mass inflation or the dynamical stability of the black hole interior~\cite{FrolovZelnikov:2026QTGMassInflation,DiFilippo:2026QTGMassInflation,Liu:2026ltw}. We therefore also ask whether the globally regular Born-Infeld charged black holes can support a degenerate inner horizon while maintaining a simple outer event horizon.

In this work, we construct exact static, spherically symmetric charged solutions in this four-dimensional NPQTG-Born-Infeld model. We show that Born-Infeld nonlinearities modify the radial dependence of the gravitational field equation so that, within a finite parameter region, the charged geometry avoids the finite-radius singularity of the vacuum branch and extends from an asymptotically flat exterior to a finite-curvature AdS-type core. The resulting regular geometry violates both the timelike and null convergence conditions in a neighborhood of the core, while these conditions are satisfied asymptotically. We further find that the globally regular sector contains both horizonless configurations and RBHs, separated by a degenerate-horizon boundary, and that the RBHs occupy a parameter region of nonzero width. Within this region, a continuous branch exists with a triple-degenerate inner horizon and a simple outer event horizon satisfying $\kappa_-=0$ and $\kappa_+\neq0$. Here nonlinear matter instead restores global regularity to a gravitational branch that is singular in vacuum.

The remainder of this paper is organized as follows. Section~\ref{sec2} presents the spherically symmetric reduction of QTG coupled to NED and its specialization to the four-dimensional NPQTG-Born-Infeld theory. In Section~\ref{sec3}, we analyze the asymptotic behavior, central geometry, global regularity, and geometric convergence conditions. In Section~\ref{sec4}, we investigate the horizon structure of the globally regular solutions and the branch with a triple-degenerate inner horizon. Finally, the conclusions and discussion are presented in Section~\ref{sec5}.

\section{Quasi-topological gravity coupled to Born-Infeld electrodynamics}
\label{sec2}

\subsection{Spherically symmetric dynamics and a four-dimensional covariant realization}
\label{sec2-1}
We work in the single-function static, spherically symmetric sector of four-dimensional NPQTG~\cite{Borissova:2026Regular4D}. The total action is
\begin{align}
\label{eq1}
S[g,\mathcal {A}]=S_{\rm grav}[g]+S_{\rm BI}[g,{\mathcal A}].
\end{align}

A curvature-only four-dimensional covariant realization of the spherical sector considered here is provided by the non-polynomial quasi-topological construction of Ref.~\cite{Borissova:2026Regular4D},
\begin{align}
\label{eq2}
S_{\rm grav}& =\frac{1}{16\pi G}\int {\rm d}^{4}x\sqrt{-g}\left[H_{2}(\mathcal P)-H_{3}(\mathcal P)\mathcal H \right.
\nonumber\\
&\left. +H_{4}(\mathcal P)\mathcal K +2H_{4}'(\mathcal P)
\left(\mathcal H^{2}-\mathcal T\right)
\right],
\end{align}
where $\mathcal P$, $\mathcal H$, $\mathcal K$, and $\mathcal T$ are four-dimensional covariant curvature combinations~\cite{Borissova:2026Regular4D}. For the single-function subclass, the three functions $H_i$ are determined by a characteristic function $H\left(p\right)$ through
\begin{subequations}
\label{eq2a}
\begin{align}
H_{3}\left(p\right)&=\frac{2}{3}H'\left(p\right),
\\
H_{2}\left(p\right)&=H\left(p\right)-\frac{2}{3}pH'\left(p\right),
\\
H_{4}\left(p\right)&=-\frac{p}{6}\int\frac{H'\left(p\right)}{p^{2}}{\rm d}p.
\end{align}
\end{subequations}

The Born-Infeld sector is described by
\begin{align}
\label{eq3}
S_{\rm BI}=\int {\rm d}^{4}x\sqrt{-g}\,L(\mathcal F),
\end{align}
where $L(\mathcal F)=b^{2}\left(1-\sqrt{1-\mathcal F/b^{2}}\right)$ is the Lagrangian density of the Born–Infeld nonlinear electromagnetic field with $\mathcal F=-F_{\mu\nu}F^{\mu\nu}/2$, $F_{\mu\nu}=\partial_{\mu}\mathcal A_{\nu}-\partial_{\nu}\mathcal A_{\mu}$, and the parameter $b$ sets the characteristic field-strength scale of the Born-Infeld nonlinearity. In the weak-field regime $|\mathcal F|\ll b^{2}$, $L(\mathcal F)=\mathcal F/2+\mathcal O(\mathcal F^{2}/b^{2})$, recovering Maxwell electrodynamics in the normalization adopted here.

To obtain the reduced spherical field equations, it is convenient to first use the general $D$-dimensional QTG-NED reduction and then specialize the result to four dimensions. For a static, spherically symmetric charged configuration, the $D$-dimensional ansatz is given by
\begin{align}
\label{eq4}
{\rm d}s^{2}=-N^{2}\left(r\right)f\left(r\right){\rm d}t^{2}+\frac{{\rm d}r^{2}}{f\left(r\right)}+r^{2}{\rm d}\Omega_{D-2}^{2},
\end{align}
with ${\mathcal A}_{\mu}{\rm d}x^{\mu}=\phi\left(r\right){\rm d}t$ and $\mathcal E\left(r\right)=\phi'\left(r\right)$. The electromagnetic invariant becomes ${\cal F}={\mathcal E^{2}}/{N^{2}}$ and therefore contains no explicit dependence on $f\left(r\right)$. The spherical gravitational dynamics can be expressed in terms of
\begin{align}
\label{eq5}
p\left(r\right)=\frac{1-f\left(r\right)}{r^{2}}
\end{align}
and a single-variable characteristic function~\cite{Borissova:2026Regular4D,PinedoSoto:2026QTGBI}. In the four-dimensional NPQTG convention, this function is denoted by $H\left(p\right)$, with $H\left(p\right)=6p$ corresponding to Einstein gravity, whereas the general QTG-NED reduction uses the normalization $h\left(p\right)=p$ in the Einstein limit~\cite{PinedoSoto:2026QTGBI}. The two conventions are related by
\begin{align}
\label{eq6}
H\left(p\right)=6h\left(p\right).
\end{align}

Within the single-function NPQTG sector, the characteristic function specifies the higher-curvature theory and need not be restricted to the Hayward  form~\cite{Borissova:2026Regular4D}. The different choices of this function correspond to different covariant gravitational models. The Hayward-type regular vacuum geometry is associated, in the normalization adopted here, with $h_{\rm H}\left(p\right)=p/(1-\ell^{2}p)$. We consider instead
\begin{align}
\label{eq7}
h\left(p\right)=\frac{p}{1+\ell^{2}p},
\end{align}
where $\ell$ sets the characteristic length scale of the higher-curvature corrections. The two characteristic functions share the same Einstein limit at low curvature but correspond to different higher-curvature corrections. In particular, $h\left(p\right)=p-\ell^{2}p^{2} +\ell^{4}p^{3}-\ell^{6}p^{4}+\cdots$.  Thus, Eq.~(\ref{eq7}) specifies a distinct NPQTG model rather than a charged deformation of the Hayward branch.

For Eq.~(\ref{eq7}), $H\left(p\right)=6p/(1+\ell^{2}p)$, and Eq.~(\ref{eq2a}) gives
\begin{subequations}
\label{eq7H}
\begin{align}
H_{2}\left(p\right)
&=
\frac{2p\left(1+3\ell^{2}p\right)}
{\left(1+\ell^{2}p\right)^{2}},
\\
H_{3}\left(p\right)
&=
\frac{4}
{\left(1+\ell^{2}p\right)^{2}},
\\
H_{4}\left(p\right)
&=
1+\frac{\ell^{2}p}{1+\ell^{2}p}
-2\ell^{2}p
\ln\left|
\frac{1+\ell^{2}p}{\ell^{2}p}
\right|.
\end{align}
\end{subequations}
The integration freedom in $H_{4}$ has been fixed for convenience. Other choices differ by a term linear in $p$ and leave the spherical field equations unchanged. The behavior of these coefficient functions along the globally regular solutions is examined in Appendix~\ref{app1}.

In the normalization adopted above, the reduced action takes the form~\cite{PinedoSoto:2026QTGBI}
\begin{align}
\label{eq8}
S_{\rm red}&=\frac{\Omega_{D-2}}{16\pi G_D}\int {\rm d}t {\rm d}r[
-(D-2)N'\left(r\right)r^{D-1}h\left(p\right)
\nonumber\\
&+16\pi G_D N\left(r\right)r^{D-2}L(\mathcal F)],
\end{align}
where $\Omega_{D-2}={2\pi^{(D-1)/2}}/{\Gamma[(D-1)/2]}$ is the area of the unit $(D-2)$-sphere.

Variation of Eq.~(\ref{eq8}) with respect to $N\left(r\right)$, $f\left(r\right)$, and $\phi\left(r\right)$ gives
\begin{subequations}
\begin{align}
\label{eq10}
&N'\left(r\right)=0,
\\
&\left[
r^{D-2}\frac{\mathcal E}{N}
\frac{\partial L}{\partial\mathcal F}
\right]'=0,
\\
&\left[r^{D-1}h\left(p\right)\right]'
+\frac{16\pi G_D}{D-2}
r^{D-2}
\left(
L-2\mathcal F
\frac{\partial L}{\partial\mathcal F}
\right)=0.
\end{align}
\end{subequations}
The first equation allows $N=1$ to be chosen by rescaling the time coordinate, so that ${\mathcal F}=\mathcal E^{2}$. Defining
$\mathcal L(\mathcal E)=L(\mathcal F)|_{N=1}$
and
$\mathcal D(\mathcal E)={\rm d}\mathcal L/{\rm d}\mathcal E$,
the electromagnetic equation integrates to
$r^{D-2}\mathcal D=Q$. The gravitational equation can then be written as
\begin{align}
\label{eq11}
\left[r^{D-1}h\left(p\right)\right]'=-U\left(r\right),
\end{align}
where $U\left(r\right)={16\pi G_D}\left[r^{D-2}\mathcal L(\mathcal E)-Q\mathcal E \right]/({D-2})$. 

We now specialize to $D=4$. For Born-Infeld electrodynamics,
\begin{align}
\label{eq12}
\mathcal L_{\rm BI}(\mathcal E)
=b^{2}
\left(
1-\sqrt{1-\frac{\mathcal E^{2}}{b^{2}}}
\right),
\end{align}
with $\mathcal D(\mathcal E)
={\mathcal E}/{\sqrt{1-\mathcal E^{2}/b^{2}}}$. The electric equation therefore becomes
\begin{align}
\label{eq13}
r^{2}\frac{\mathcal E}
{\sqrt{1-\mathcal E^{2}/b^{2}}}
=Q.
\end{align}
For the characteristic function in Eq.~(\ref{eq7}), the gravitational equation reduces to
\begin{align}
\label{eq14}
\frac{\rm d}{{\rm d}r}\left[r^{3}\frac{p}{1+\ell^{2}p}
\right]=-8\pi G \left[r^{2}\mathcal L_{\rm BI}(\mathcal E) -Q\mathcal E
\right].
\end{align}
Eqs.~(\ref{eq13}) and~(\ref{eq14}) determine the electric field and the curvature variable $p\left(r\right)$. The metric function then follows from $f\left(r\right)=1-r^{2}p\left(r\right)$.

\subsection{Charged black hole solution}
\label{sec2-2}
Eq.~(\ref{eq13}) directly gives the four-dimensional Born-Infeld electric field
\begin{align}
\label{eq15}
\mathcal E\left(r\right) = \frac{Q}{\sqrt{r^{4}+Q^{2}/b^{2}}}.
\end{align}
Unlike the Maxwell electric field, Eq.~(\ref{eq15}) remains finite as $r\to0$ and recovers the Coulomb behavior $\mathcal E\left(r\right)\sim Q/r^{2}$ at large radii. The sign of the electric field changes with the charge $Q$, whereas the gravitational equation depends only on the magnitude of the charge. Therefore, $Q>0$ is assumed in the following without loss of generality. Substituting Eq.~(\ref{eq15}) into $U\left(r\right)$ gives
\begin{align}
\label{eq16}
U\left(r\right)=-8\pi G b^{2}\left(\sqrt{r^{4}+\frac{Q^{2}}{b^{2}}}-r^{2}\right).
\end{align}
Substituting Eq.~(\ref{eq16}) into Eq.~(\ref{eq11}) and integrating, the gravitational field equation can be written as
\begin{align}
\label{eq17}
h\left(p\right) = \frac{\mu+Z\left(r\right)}{r^{3}},
\quad
Z\left(r\right) = \int_{r}^{\infty}U(r'){\rm d}r',
\end{align}
where the integration constant $\mu=2GM$ is fixed by the asymptotic mass.

It is convenient to introduce the dimensionless radial variable
\begin{align}
\label{eq18}
\rho=\frac{br^{2}}{Q}.
\end{align}
For the integration variable $r'$ in $Z\left(r\right)$, we similarly set $x=b{r'}^{2}/Q$, so that ${\rm d}r'=\frac{1}{2}(Q/b)^{1/2}x^{-1/2}{\rm d}x$. As $r'$ runs from $r$ to infinity, $x$ runs from $\rho$ to infinity. Using Eq.~(\ref{eq16}), one obtains
\begin{align}
\label{eq19}
Z\left(r\right) = -4\pi G\sqrt{b}\,Q^{3/2}J_{4}\left(\rho\right),
\end{align}
where
\begin{align}
\label{eq20}
J_{4}\left(\rho\right) = \int_{\rho}^{\infty}
x^{-1/2}\left(\sqrt{x^{2}+1}-x\right)
{\rm d}x.
\end{align}
The integral in Eq.~(\ref{eq20}) can also be expressed analytically in terms of the Gauss hypergeometric function,
\begin{align}
\label{eq21}
J_{4}\left(\rho\right)=\frac{2}{3}\rho^{3/2}\left[1-F_{4}\left(\rho\right)\right],
\end{align}
where $F_{4}\left(\rho\right)={}_2F_{1}\left(-\frac{1}{2},-\frac{3}{4};\frac{1}{4};-\frac{1}{\rho^{2}}\right)$.

For later convenience, define $H_{\rm BI}\left(r\right)=[\mu+Z\left(r\right)]/r^3$. The gravitational field equation then takes the algebraic form
\begin{align}
\label{eq20+}
h\left(p\right)=H_{\rm BI}\left(r\right),
\end{align}
where $H_{\rm BI}\left(r\right)$ contains both the mass term and the contribution from the Born-Infeld electromagnetic field. Once $H_{\rm BI}\left(r\right)$ is known, the characteristic relation determines $p\left(r\right)$ and hence the metric function. Using this definition together with $Z\left(r\right)$, we obtain
\begin{align}
\label{eq21+}
H_{\rm BI}\left(r\right)&=\frac{\mu}{r^{3}}-\frac{4 \pi G\sqrt{b}Q^{3/2}}{r^{3}}J_{4}\left(\rho\right)
\nonumber\\
&=\frac{2GM}{r^{3}}+\frac{8 \pi G b^{2}}{3}\left[{}_2F_{1}\left(-\frac{1}{2},-\frac{3}{4};\frac{1}{4};-\frac{1}{\rho^{2}}
\right)-1\right].
\end{align}
Eq.~(\ref{eq21+}) gives the integral and hypergeometric representations of $H_{\rm BI}\left(r\right)$ and agrees with the four-dimensional QTG-Born-Infeld result of Ref.~\cite{PinedoSoto:2026QTGBI}. Combining Eqs.~(\ref{eq7}) and~(\ref{eq20+}) gives
\begin{align}
\label{eq22+}
\frac{p}{1+\ell^{2}p}=H_{\rm BI}\left(r\right).
\end{align}
Solving Eq.~(\ref{eq22+}) gives
\begin{align}
\label{eq23+}
p\left(r\right)=\frac{H_{\rm BI}\left(r\right)}{1-\ell^{2}H_{\rm BI}\left(r\right)}.
\end{align}
Finally, using Eq.~(\ref{eq5}), the metric function is
\begin{align}
\label{eq24+}
f\left(r\right)=1-\frac{r^{2}H_{\rm BI}\left(r\right)}{1-\ell^{2}H_{\rm BI}\left(r\right)}.
\end{align}
Eqs.~(\ref{eq23+}) and~(\ref{eq24+}) give the exact static, spherically symmetric charged solution,  characterized by the parameters $M$, $Q$, $b$, and $\ell$. Its global extension depends on whether $H_{\rm BI}\left(r\right)$ reaches the critical value $1/\ell^{2}$ at a finite radius, where the characteristic relation in Eq.~(\ref{eq23+}) becomes singular.

\section{Regularity and spacetime geometry}
\label{sec3}
Geometric regularity of the exact solution in Eq.~(\ref{eq24+}) must be checked over the full radial domain.\footnote{Throughout this work, regularity refers to geometric regularity of the static, spherically symmetric spacetime, namely the absence of curvature singularities. The action-level behavior of the globally regular solutions is discussed separately in Appendix~\ref{app1}.} In the present non-polynomial model, the metric function depends on $H_{\rm BI}\left(r\right)$ through the characteristic relation in Eq.~(\ref{eq23+}). Regular behavior near the center alone is therefore insufficient, since an additional singularity may occur at a finite radius. We examine below the asymptotic behavior, the central geometry, the condition that excludes such finite-radius singularities, and the geometric convergence conditions of the resulting regular solutions.

\subsection{Asymptotic behavior and the finite-radius vacuum singularity}
\label{sec3-1}
At large radius, Eq.~(\ref{eq21+}) gives
\begin{align}
\label{eq26}
H_{\rm BI}\left(r\right)=\frac{2GM}{r^{3}}-\frac{4\pi G Q^{2}}{r^{4}}+\frac{ \pi G Q^{4}}{5b^{2}r^{8}}+\mathcal O(r^{-12}).
\end{align}
The Born-Infeld corrections enter  beyond the Maxwell term in this expansion. In the weak-curvature regime $\ell^2H_{\rm BI}\left(r\right)\ll1$, 
$p=H_{\rm BI}+\ell^{2}H_{\rm BI}^{2}+\cdots$, and the metric function becomes
\begin{align}
\label{eq27}
f\left(r\right)\!=\!\!1-\!\frac{2GM}{r}\!+\!\frac{4 \pi G Q^{2}}{r^{2}}\!-\!\frac{4G^{2}M^{2}\ell^{2}}{r^{4}}\!+\mathcal O(r^{-5}).
\end{align}
The spacetime is asymptotically flat. Its leading mass and charge terms agree with those of four-dimensional Einstein-Maxwell theory, while the NPQTG corrections appear only at higher orders. In the limit $\ell\to0$, Eq.~(\ref{eq24+}) reduces to the corresponding Einstein-Born-Infeld solution, providing a direct consistency check of the exact solution~\cite{Garcia:1984BornInfeld,PinedoSoto:2026QTGBI}.

The strong-curvature behavior is controlled by the characteristic relation in Eq.~(\ref{eq23+}). In the vacuum limit $Q=0$, 
\begin{align}
\label{eq28}
H_{0}\left(r\right)=\frac{2GM}{r^{3}}, \quad p\left(r\right)=\frac{H_{0}\left(r\right)}{1-\ell^{2}H_{0}\left(r\right)}.
\end{align}
As the radius decreases, $H_{0}\left(r\right)$ grows monotonically from zero and reaches $H_{0}(r_{\rm s})=1/\ell^{2}$ at $r_{\rm s}=(2GM\ell^{2})^{1/3}$. At this radius $p\left(r\right)$ diverges, and the vacuum branch cannot be continued regularly toward the center. The sign choice in Eq.~(\ref{eq7}) therefore differs qualitatively from the Hayward-type branch and produces a finite-radius branch singularity in vacuum. A finite Born-Infeld electric field at $r=0$ is not sufficient by itself to remove this obstruction. What matters is whether the full radial profile of $H_{\rm BI}\left(r\right)$ remains below the critical value $1/\ell^{2}$ at which the characteristic relation becomes singular.

\subsection{Global regularity condition}
\label{sec3-2}
To analyze $H_{\rm BI}\left(r\right)$ over the full radial domain, we introduce
\begin{align}
\label{eq30}
\hat h\!=\!\ell^{2}H_{\rm BI},
\hat p\!=\!\ell^{2}p,
\beta \!=\! \frac{8 \pi G b^{2}\ell^{2}}{3},
A \!=\! \ell^{2}\mu \left( \frac{b}{Q} \right)^{3/2}.
\end{align}
In terms of these variables, $\hat h\left(\rho\right)$ is
\begin{align}
\label{eq31}
\hat h\left(\rho\right)=\rho^{-3/2}\left[A-\frac{3}{2}\beta J_{4}\left(\rho\right)\right].
\end{align}
Here $J_{4}\left(\rho\right)$ is positive and monotonically decreasing, with $J_{4}(\infty)=0$. Its central value is $J_{4}^{(0)}=J_{4}(0)\simeq 2.47210$~\cite{PinedoSoto:2026QTGBI}.

On the other hand, Eq.~(\ref{eq23+}) can be written in dimensionless form as
\begin{align}
\label{eq32}
\hat{h}=\frac{\hat{p}}{1+\hat{p}},
\quad
\hat{p}=\frac{\hat{h}}{1-\hat{h}}.
\end{align}
The crossing $\hat h=0$ is geometrically regular, since it corresponds simply to $\hat p=0$. The obstruction occurs instead at $\hat h=1$, where $\hat p$ diverges.

To determine whether this critical value can be reached, we first examine the limiting behavior of $\hat h\left(\rho\right)$. From Eq.~(\ref{eq31}), since $J_{4}\left(\rho\right)\to J_{4}^{(0)}$ as $\rho\to0$, the condition
\begin{align}
\label{eq34}
A<\frac{3}{2}\beta J_{4}^{(0)}
\end{align}
implies $\hat{h}\left(\rho\right)\to-\infty$ as $\rho\to0$. At large $\rho$, by contrast, $\hat{h}\left(\rho\right)\to0^{+}$. Hence, any solution satisfying Eq.~(\ref{eq34}) crosses $\hat h=0$ at a finite radius and reaches a positive maximum before approaching zero from above at infinity. Since $\hat h\left(\rho\right)$ is smooth for every finite $\rho>0$, global regularity requires this maximum to remain below $1$.

From the definition of $J_{4}\left(\rho\right)$,
\begin{align}
\label{eq35+}
\frac{{\rm d}J_{4}}{{\rm d}\rho}=-\rho^{-1/2}\left(\sqrt{1+\rho^{2}}-\rho\right).
\end{align}
Let $\rho_{\rm m}$ denote the position of the positive maximum of $\hat{h}\left(\rho\right)$. The condition ${\rm d}\hat{h}/{\rm d}\rho=0$ then gives
\begin{align}
\label{eq35}
A-\frac{3}{2}\beta J_{4}(\rho_{\rm m})=\beta\sqrt{\rho_{\rm m}}\left(\sqrt{1+\rho_{\rm m}^{2}}-\rho_{\rm m}\right).
\end{align}
Using this relation, the maximum value of $\hat h$ can be written as
\begin{align}
\label{eq36}
\hat{h}_{\max}=\beta\frac{\sqrt{1+\rho_{\rm m}^{2}}-\rho_{\rm m}}{\rho_{\rm m}}.
\end{align}
Global regularity requires $\hat h_{\max}<1$, and the boundary of the regular region is reached when 
\begin{align}
\label{eq37}
\hat{h}_{\max}=1.
\end{align}
Combining Eqs.~(\ref{eq35})-(\ref{eq37}), and using $\rho_{\rm m}=\rho_{\rm c}$, the regularity boundary is follows
\begin{align}
\label{eq39}
A_{\rm reg}\left(\beta\right)=\frac{3}{2}\beta J_{4}(\rho_{\rm c})+\rho_{\rm c}^{3/2},
\quad
\rho_{\rm c}=\frac{\beta}{\sqrt{1+2\beta}}.
\end{align}
Accordingly, the globally regular branch is defined by
\begin{align}
\label{eq40}
A<A_{\rm reg}\left(\beta\right).
\end{align}
For $\beta>0$, one also has
\begin{align}
\label{eq41}
A_{\rm reg}\left(\beta\right)<\frac{3}{2}\beta J_{4}^{(0)}.
\end{align}
Together with Eq.~(\ref{eq40}), this implies Eq.~(\ref{eq34}), so every globally regular solution belongs to the branch with $\hat h\to-\infty$ at the center. Moving outward, $\hat h\left(\rho\right)$ crosses zero, reaches a positive maximum below $1$, and approaches zero from above at infinity.
\begin{figure}[t]
\centering
\includegraphics[width=1.0\linewidth]{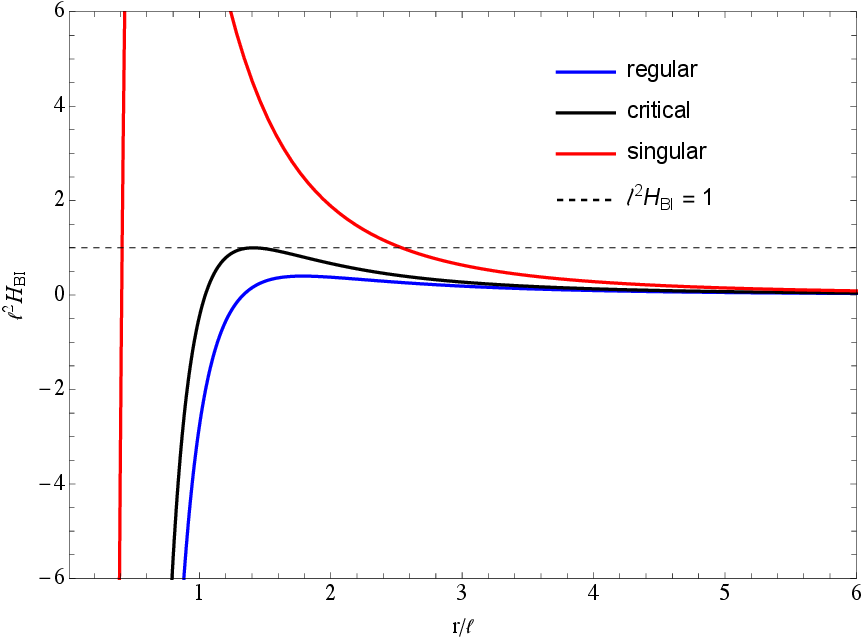}
\caption{Dimensionless function $\ell^{2}H_{\rm BI}$ as a function of $r/\ell$, for $G=b=\ell=Q=1$, corresponding to $\beta=8\pi/3$. The blue, black, and red curves correspond to $A=0.8A_{\rm reg}\simeq9.28797$, $A=A_{\rm reg}\simeq 11.60996$, and $A\simeq 21.33764$, respectively. The gray dashed line marks the critical value $\ell^{2}H_{\rm BI}=1$.}
\label{fig1}
\end{figure}

Eqs.~(\ref{eq39})-(\ref{eq41}) show how the Born-Infeld field changes the global structure of the vacuum-singular branch. In vacuum, $H_{0}\left(r\right)=\mu/r^{3}$ inevitably reaches the critical value $1/\ell^{2}$ at a finite radius. The Born-Infeld contribution changes this monotonic profile, as illustrated in  Fig.~\ref{fig1}. For $A<A_{\rm reg}\left(\beta\right)$, the maximum of $\ell^{2}H_{\rm BI}$ remains below unity and the finite-radius singularity is avoided. At $A=A_{\rm reg}$, the maximum reaches the critical value, whereas for $A>A_{\rm reg}$ it exceeds this value and the solution becomes singular. The regularity condition in Eq.~(\ref{eq40}) therefore identifies a finite parameter region in which the vacuum-singular branch is replaced by globally regular charged geometries.

\subsection{Central geometry and regularity}
\label{sec3-3}
Within the globally regular region, Eq.~(\ref{eq19}) shows that $Z\left(r\right)$ approaches a finite limit as $r\to0$. We define 
\begin{align}
\label{eq42}
\Delta = \mu-4\pi G\sqrt{b}Q^{3/2}J_{4}^{(0)}.
\end{align}
The condition $A<3\beta J_{4}^{(0)}/2$ is equivalent to $\Delta<0$. Since 
$A<A_{\rm reg}\left(\beta\right)$ implies this inequality, every globally regular solution lies on the $\Delta<0$ branch. Near the center, Eq.~(\ref{eq42}) gives
\begin{align}
\label{eq43}
H_{\rm BI}\left(r\right)=\frac{\Delta}{r^{3}}+\frac{8\pi G bQ}{r^{2}}-\frac{8\pi G b^{2}}{3}+\mathcal O(r^{2}).
\end{align}
Since $\Delta<0$, $H_{\rm BI}\left(r\right)\to-\infty$ near the center. Substituting this asymptotic behavior into Eq.~(\ref{eq23+}) gives
\begin{align}
\label{eq44}
p\left(r\right)=-\frac{1}{\ell^{2}}-\frac{r^{3}}{\ell^{4}\Delta}
+\mathcal O(r^{4}).
\end{align}
\begin{figure*}[t]
\centering
\begin{subfigure}[t]{0.45\textwidth}
\centering
\includegraphics[width=\linewidth]{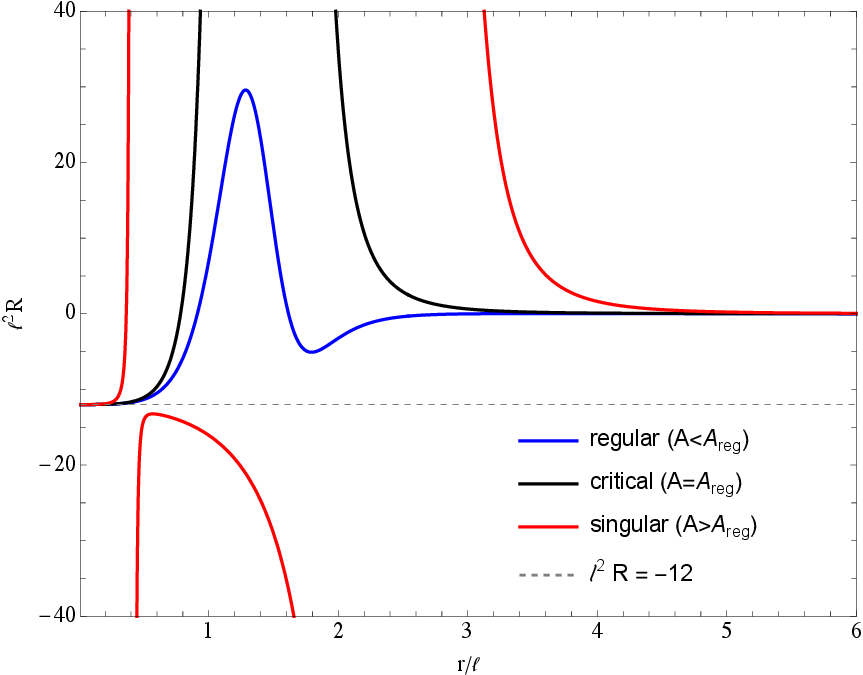}
\caption{}
\label{fig2a}
\end{subfigure}
\hfill
\begin{subfigure}[t]{0.45\textwidth}
\centering
\includegraphics[width=\linewidth]{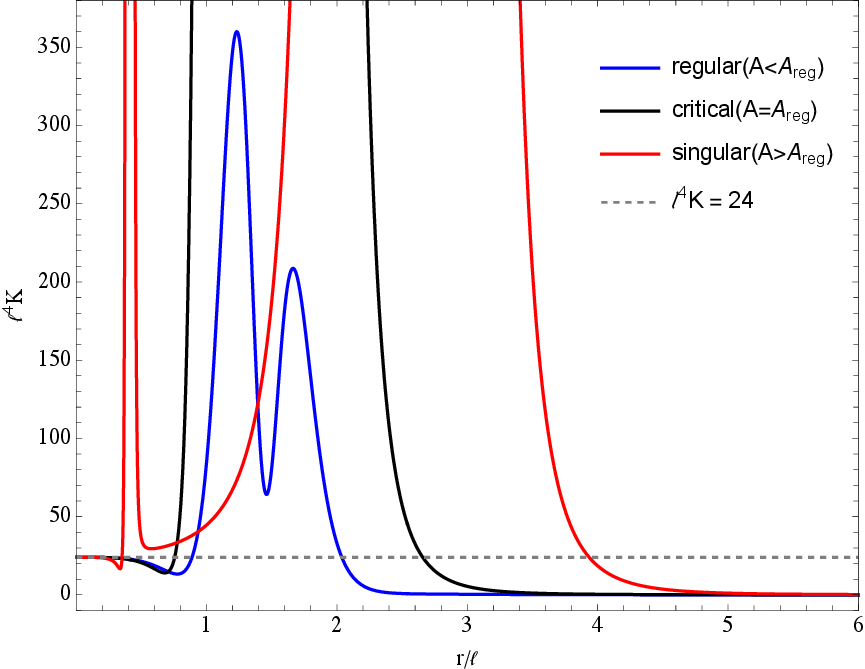}
\caption{}
\label{fig2b}
\end{subfigure}
\caption{Dimensionless curvature invariants for representative regular, critical, and singular solutions. (a) Dimensionless Ricci scalar $\ell^2 R$ as a function of $r/\ell$. (b) Dimensionless Kretschmann scalar $\ell^4 K$ as a function of $r/\ell$. Both panels use $G=b=\ell=Q=1$, corresponding to $\beta=8\pi/3$. The blue, black, and red curves correspond to $A\simeq9.28797$, $A=A_{\rm reg}\simeq11.60996$, and $A\simeq21.33764$, respectively.}
\label{fig2}
\end{figure*}
The metric function therefore has the near-center expansion
\begin{align}
\label{eq45}
f\left(r\right)=1+\frac{r^{2}}{\ell^{2}}+\frac{r^{5}}{\ell^{4}\Delta}+\mathcal O(r^{6}).
\end{align}
The leading behavior $f\left(r\right)=1+r^{2}/\ell^{2}+\cdots$
describes an AdS-type core with curvature scale $\ell$. This regular geometry does not imply that the Born-Infeld stress tensor is itself regular at the center. For the purely electric configuration,
\begin{align}
\label{eq45+}
T^{t}_{t}=T^{r}_{r}
=L-2\mathcal F L_{\mathcal F}
=-\rho_{\rm BI},
\quad
T^{\theta}_{\theta}=T^{\phi}_{\phi}=L,
\end{align}
where $\rho_{\rm BI} =2\mathcal F L_{\mathcal F}-L =b^{2}\left(
\sqrt{1+\frac{Q^{2}}{b^{2}r^{4}}}-1
\right)$.
Using Eq.~(\ref{eq15}), the near-center behavior is
\begin{align}
\label{eq46+}
\rho_{\rm BI}
=\frac{bQ}{r^{2}}-b^{2}+\mathcal O(r^{2}),
\quad
L\to b^{2}.
\end{align}
Although the Born-Infeld electric  field remains finite as $r\to0$, the local energy density and radial  pressure diverge as $r^{-2}$. The divergence is integrable because
\begin{align}
\label{eq47+}
r^{2}\rho_{\rm BI}\left(r\right)
=bQ-b^{2}r^{2}+\mathcal O(r^{4}).
\end{align}
The combination entering the reduced gravitational equation nevertheless remains finite,
\begin{align}
\label{eq48+}
U\left(r\right)=-8\pi G r^{2}\rho_{\rm BI}\left(r\right)=-8\pi G bQ+\mathcal O(r^{2}).
\end{align}
The regular AdS-type core therefore does not result from regularity of the matter stress tensor itself, but from the combined effect of the Born-Infeld matter contribution and the non-polynomial gravitational dynamics.

The regularity of the center can also be verified directly from curvature invariants. For the static, spherically symmetric metric, the Ricci scalar and the Kretschmann scalar are
\begin{subequations}
\label{eq46}
\begin{align}
R\left(r\right)&=-f''\left(r\right)-\frac{4f'\left(r\right)}{r}+\frac{2[1-f\left(r\right)]}{r^{2}},
\\
K\left(r\right)&=[f''\left(r\right)]^{2}+4\left[\frac{f'\left(r\right)}{r}\right]^{2}+4\left[\frac{1-f\left(r\right)}{r^{2}}\right]^{2}.
\end{align}
\end{subequations}
The explicit expressions of $R\left(r\right)$ and $K\left(r\right)$ are given in   Appendix~\ref{app-0}.  At $r=0$, the curvature invariants approach
\begin{align}
\label{eq49}
R(0)=-\frac{12}{\ell^{2}},
\quad
K(0)=\frac{24}{\ell^{4}}.
\end{align}
Both curvature invariants remain finite at the center, in agreement with the AdS-type behavior in Eq.~(\ref{eq45}). Although the covariant coefficient functions become singular as $p\to-1/\ell^{2}$, the corresponding spherically reduced action remains locally integrable, as discussed in Appendix~\ref{app1}. For every finite $r>0$, $H_{\rm BI}\left(r\right)$ and its radial derivatives are finite. A finite-radius curvature singularity can therefore arise only when $1-\ell^{2}H_{\rm BI}\left(r\right)=0$. At such a point $p\left(r\right)$ diverges, and the term $4(1-f)^{2}/r^{4}=4p^{2}$ in the Kretschmann scalar diverges with it. The global condition in Eq.~(\ref{eq40}) excludes precisely this possibility.

Fig.~\ref{fig2} provides a further verification of the above analysis. For $A<A_{\rm reg}$, both curvature invariants represented by the blue curves evolve continuously from finite values at the center and approach zero in the asymptotic region. For $A>A_{\rm reg}$, both invariants represented by the red curves diverge at a finite radius, while $A=A_{\rm reg}$, shown by the black curves, corresponds to the critical case separating these two behaviors. Thus, Eq.~(\ref{eq40}) not only guarantees regularity at the center, but also excludes additional curvature singularities associated with the singular characteristic relation throughout the finite radial domain.

\subsection{Geometric convergence conditions}
\label{sec3-4}

In higher-curvature gravity, the matter energy conditions do not in general determine the geometric convergence conditions entering the singularity theorems. The latter should instead be evaluated directly on the solution~\cite{Borissova:2026prd}. For the static, single-function spherically symmetric metric considered here, the relevant geometric inequalities can be written as
\begin{subequations}
\label{eqa1}
\begin{align}
\mathcal C_{1}\left(r\right)&=\frac{f'\left(r\right)}{r}
+\frac{1}{2}f''\left(r\right)\geq0,
\\
\mathcal C_{2}\left(r\right)&=\frac{1-f\left(r\right)}{r^{2}}+\frac{1}{2}f''\left(r\right)\geq0.
\end{align}
\end{subequations}
The timelike convergence condition (TCC) requires both inequalities in Eq.~(\ref{eqa1}) to hold, whereas the nontrivial null convergence condition (NCC) reduces to the second inequality, with the radial null condition identically saturated.

Using Eqs.~(\ref{eq44}) and~(\ref{eq45}), their near-center behavior is
\begin{subequations}
\label{eqa2}
\begin{align}
\mathcal C_{1}\left(r\right)&=\frac{3}{\ell^{2}}+\frac{15r^{3}}{\ell^{4}\Delta}+\mathcal O(r^{4}),
\\
\mathcal C_{2}\left(r\right)&=\frac{9r^{3}}{\ell^{4}\Delta}+\mathcal O(r^{4}).
\end{align}
\end{subequations}
Since every globally regular solution satisfies $\Delta<0$, one has $\mathcal C_{1}(0)=3/\ell^{2}>0$ and $\mathcal C_{2}(0)=0$. Thus, at the exact center the TCC is satisfied and the NCC is saturated. For sufficiently small but nonzero $r$, however, $\mathcal C_{2}\left(r\right)<0$. Therefore, both the TCC and the NCC are violated in a neighborhood of the regular core.

At large radius, Eq.~(\ref{eq27}) gives
\begin{subequations}
\label{eqa3}
\begin{align}
\mathcal C_{1}\left(r\right)&=\frac{4\pi GQ^{2}}{r^{4}}+\mathcal O(r^{-6}),
\\
\mathcal C_{2}\left(r\right)&=\frac{8\pi GQ^{2}}{r^{4}}+\mathcal O(r^{-6}).
\end{align}
\end{subequations}
Both the TCC and the NCC are therefore satisfied asymptotically. The globally regular branch thus violates the geometric convergence conditions in a neighborhood of the core while recovering them at large distances. These results follow directly from the on-shell geometry and should not be identified with energy conditions imposed on the Born-Infeld stress tensor.

\section{Horizon structure of globally regular solutions}
\label{sec4}
\subsection{Horizon equation and general classification}
\label{sec4-1}
Within the globally regular region $A<A_{\rm reg}\left(\beta\right)$, whether the geometry describes a black hole is determined by the zero structure of the metric function $f\left(r\right)$~\cite{Hayward:2005gi,CarballoRubio:2018Viability}. The Killing horizons are determined by the positive roots of
\begin{align}
\label{eq50}   
f(r_h)=0.
\end{align}
Using Eq.~(\ref{eq24+}), this condition can be written equivalently as
\begin{align}
\label{eq51}
H_{\rm BI}(r_h)=\frac{1}{r_h^{2}+\ell^{2}}.
\end{align}
Eq.~(\ref{eq51}) describes horizon formation in terms of intersections between $H_{\rm BI}\left(r\right)$ and the curve $1/(r^{2}+\ell^{2})$. This is distinct from the global regularity condition of Section~\ref{sec3}, which excludes the critical value $H_{\rm BI}=1/\ell^{2}$ at which the characteristic relation becomes singular. Global regularity and horizon formation therefore impose distinct conditions on the solution.

Within the globally regular parameter region, if Eq.~(\ref{eq50}) has no positive real root, the solution describes a regular horizonless configuration. If instead there are two distinct simple positive roots,
\begin{align}
\label{eq52}
0<r_-<r_+,
\quad
f(r_-)=f(r_+)=0,
\quad
f'(r_\pm)\neq0,
\end{align}
the smaller root $r_-$ corresponds to the inner horizon, while the larger root $r_+$ corresponds to the outer event horizon. Such two-horizon causal structures are common in RBH geometries~\cite{Hayward:2005gi,CarballoRubio:2018Viability,CarballoRubio:2021InnerHorizon,DiFilippo:2022InnerHorizon}. In the maximal analytic extension, the inner horizon has the character of a Cauchy horizon~\cite{CarballoRubio:2021InnerHorizon,DiFilippo:2022InnerHorizon}.

As the parameters vary, the transition between the horizonless and two-horizon configurations occurs when the two roots merge. The critical radius $r_e$ then satisfies
\begin{align}
\label{eq53}
f(r_e)=0,
\quad
f'(r_e)=0.
\end{align}
This configuration corresponds to a degenerate Killing horizon and defines the critical boundary between regular horizonless solutions and RBHs in the parameter space.

For the metric convention used here, the surface gravity at a Killing
horizon is
\begin{align}
\label{eq54}
\kappa_h = \frac{1}{2}\left|f'(r_h)\right|.
\end{align}
The degenerate horizon described by Eq.~(\ref{eq53}) therefore satisfies $\kappa_e=0$. For a black hole with a simple outer event horizon, on the other hand, the Hawking temperature is
~\cite{Hawking:1975vcx}
\begin{align}
\label{eq55}
T_{\rm H} = \frac{f'(r_+)}{4\pi},
\end{align}
where the physical outer horizon satisfies $f'(r_+)>0$. Hence, the degenerate horizon defined by Eq.~(\ref{eq53}) represents a zero-temperature critical state.

\subsection{Degenerate horizons and the critical boundary between horizon structures}
\label{sec4-2}
The degenerate-horizon condition in Eq.~(\ref{eq53}) defines the onset of horizon formation within the globally regular region. We now locate  this boundary in parameter space and compare it with the global regularity boundary. Similar horizon-merger conditions arise for extremal QTG-Born-Infeld black holes~\cite{PinedoSoto:2026QTGBI}.

\begin{figure*}[t]
\centering
\begin{subfigure}[t]{0.45\textwidth}
\centering
\includegraphics[width=\linewidth]{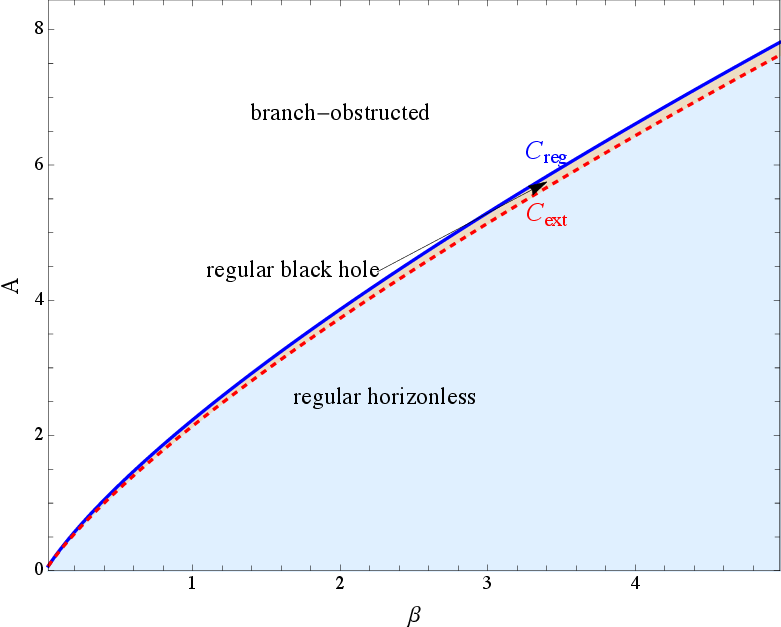}
\caption{}
\label{fig3a}
\end{subfigure}
\begin{subfigure}[t]{0.45\textwidth}
\centering
\includegraphics[width=\linewidth]{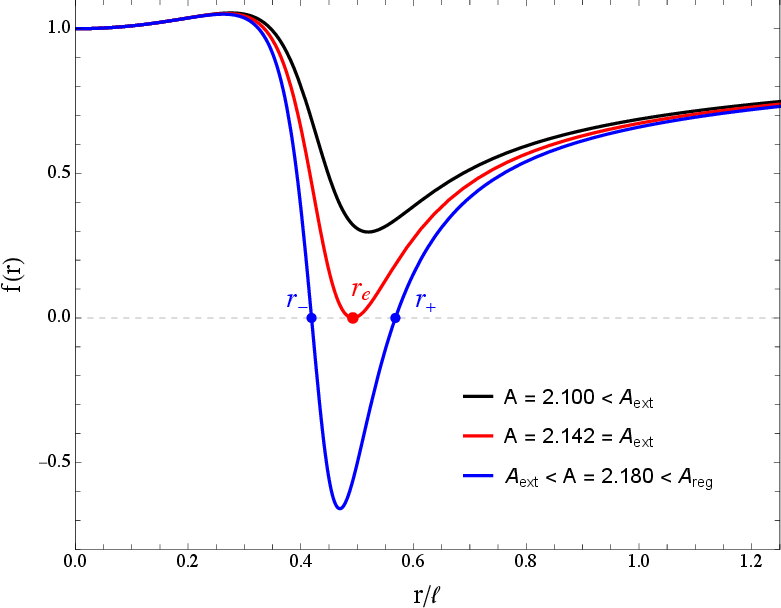}
\caption{}
\label{fig3b}
\end{subfigure}
\caption{Horizon structure of the globally regular solutions and its dependence on the model parameters. (a) Representative $(\beta,A)$ slice at $\xi=1$, bounded by the global regularity boundary $C_{\rm reg}$ and the extremal boundary $C_{\rm ext}$. (b) Representative metric functions at $(\beta,\xi)=(1,1)$. The black, red, and blue curves correspond to $A=2.10<A_{\rm ext}$, $A=A_{\rm ext}\simeq2.14232$, and $A=2.18$ respectively, with $A_{\rm ext}<A<A_{\rm reg}\simeq2.23321$.}
\label{fig3}
\end{figure*}
To express the horizon condition in a form suitable for parameter-space analysis, introduce the dimensionless combination
\begin{align}
\label{eq56}
\xi=8\pi G Q b.
\end{align}
Combining Eqs.~(\ref{eq18}) and~(\ref{eq30}) gives
\begin{align}
\label{eq57}
\left(\frac{r}{\ell}\right)^2 = \frac{\xi\rho}{3\beta}.
\end{align}
Using the horizon condition in Eq.~(\ref{eq51}) together with Eq.~(\ref{eq31}), the parameter $A$ corresponding to any horizon can be written as
\begin{align}
\label{eq58}
A=A_{\rm h}(\rho,\beta,\xi)=\frac{3}{2}\beta J_{4}\left(\rho\right)+\frac{3\beta\rho^{3/2}}
{3\beta+\xi\rho}.
\end{align}
For fixed $(\beta,\xi)$, horizons are the intersections of the horizontal line $A=\mathrm{const.}$ with $A_{\rm h}\left(\rho\right)$. No intersection corresponds to a horizonless geometry, while two intersections give the inner and outer horizons. At the transition between these cases the horizontal line is tangent to a local minimum of $A_{\rm h}\left(\rho\right)$, and the two horizons merge.

A degenerate horizon at $\rho=\rho_e$ therefore corresponds to a local minimum of $A_{\rm h}$,
\begin{align}
\label{eq60}
\left.\frac{\partial A_{\rm h}}{\partial\rho}
\right|_{\rho_e}=0,
\quad
\left.\frac{\partial^{2}A_{\rm h}}{\partial\rho^{2}}\right|_{\rho_e}>0.
\end{align}
The corresponding critical parameter is denoted by
\begin{align}
\label{eq61}
A_{\rm ext}(\beta,\xi)=A_{\rm h}(\rho_e,\beta,\xi).
\end{align}
For fixed $\xi$, Eq.~(\ref{eq61}) defines a curve $C_{\rm ext}$ in the $(\beta,A)$ plane. The global regularity boundary $A=A_{\rm reg}\left(\beta\right)$ will be denoted by $C_{\rm reg}$. The two curves encode different conditions. $C_{\rm reg}$ marks the onset of the finite radius branch obstruction, whereas $C_{\rm ext}$ marks the onset of black hole horizons within the regular region. If $A_{\rm ext}<A_{\rm reg}$, the corresponding parameter ranges are
\begin{align}
\label{eq62}
A<A_{\rm ext}
&\Longrightarrow
\text{regular horizonless},
\nonumber\\
A=A_{\rm ext}
&\Longrightarrow
\text{degenerate black hole},
\nonumber\\
A_{\rm ext}<A<A_{\rm reg}
&\Longrightarrow
\text{regular black hole}.
\end{align}
A finite RBH region exists whenever
\begin{align}
\label{eq63}
A_{\rm ext}(\beta,\xi)<A_{\rm reg}\left(\beta\right).
\end{align}
Appendix~\ref{app2} shows analytically that this inequality holds for the full physical range $\beta>0$ and $\xi>0$. The RBH solutions therefore occupy a region of nonzero width rather than a fine-tuned trajectory in parameter space. On $C_{\rm ext}$ the merged horizon has $\kappa_e=0$ and $T_{\rm H}=0$, so this curve is both the horizon-formation boundary and the zero-temperature extremal boundary.

Fig.~\ref{fig3a} shows the representative $\xi=1$ slice of the full parameter space, including the two critical boundaries and the corresponding solution regions in the $(\beta,A)$ plane. Consistent with the general result above, $C_{\rm ext}$ always lies below $C_{\rm reg}$ and further divides the globally regular parameter region into regular horizonless configurations and RBHs. The finite region between the two critical curves shows that Born-Infeld matter allows black hole horizons to form while global regularity is maintained. Fig.~\ref{fig3b} further fixes $\beta=1$ and illustrates the change in horizon structure as $A$ varies across $C_{\rm ext}$. In this case, $A_{\rm ext}\simeq2.14232$ and $A_{\rm reg}\simeq2.23321$. As $A$ increases across $C_{\rm ext}$, the initially horizonless regular spacetime first develops a degenerate horizon and then becomes a RBH with distinct simple inner and outer horizons. Thus, $C_{\rm ext}$ represents the critical state for black hole horizon formation within the globally regular sector.

\subsection{Triple-degenerate inner horizon}
\label{sec4-3}

For generic RBHs with two simple horizons, the inner Cauchy horizon has
nonzero surface gravity, and a nonvanishing $\kappa_-$ is closely
related to the exponential blueshift and mass-inflation instability near the inner  horizon~\cite{PoissonIsrael:1989,PoissonIsrael:1990,Ori:1991,CarballoRubio:2018Viability,CarballoRubio:2021InnerHorizon,DiFilippo:2022InnerHorizon}. The inner-extremal RBHs provide a different
possibility, with distinct inner and outer horizons satisfying $\kappa_-=0$ and $\kappa_+\neq0$~\cite{CarballoRubio:2022InnerExtremal,DiFilippo:2025InnerExtremal}. We examine whether this horizon structure occurs inside the RBH region obtained above.
\begin{figure*}[t]
\centering
\begin{subfigure}[t]{0.45\textwidth}
\centering
\includegraphics[width=\linewidth]{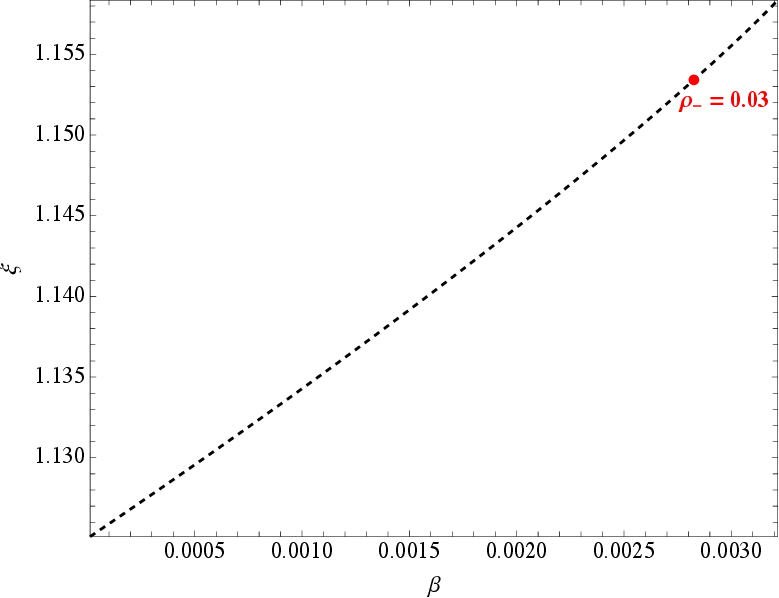}
\caption{}
\label{fig4a}
\end{subfigure}
\hfill
\begin{subfigure}[t]{0.45\textwidth}
\centering
\includegraphics[width=\linewidth]{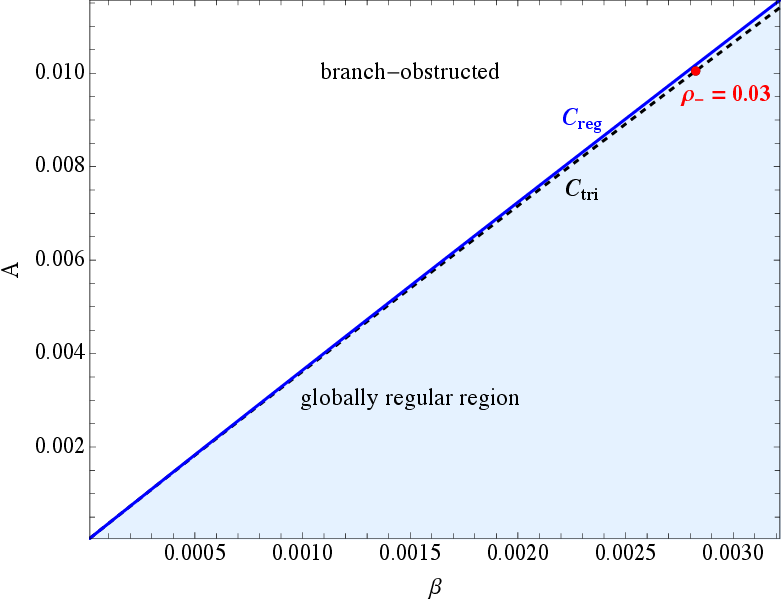}
\caption{}
\label{fig4b}
\end{subfigure}
\caption{Parameter-space structure of the triple-degenerate branch $C_{\rm tri}$. (a) Projection of the physical $\chi_-^{(-)}$ branch of $C_{\rm tri}$ onto the $(\beta,\xi)$ plane, parameterized by $0<\rho_-<\rho_-^{*}\simeq0.0421631$. The red point denotes the representative solution with $\rho_-=0.03$, $\chi_-^{(-)}\simeq4.07765$, $\beta\simeq2.82872\times10^{-3}$, $\xi\simeq1.15345$, and $A_{\rm tri}\simeq1.00574\times10^{-2}$. (b) Projection onto the $(\beta,A)$ plane. The black dashed curve denotes $C_{\rm tri}$, while the blue solid curve denotes the global regularity boundary $C_{\rm reg}$. The shaded region below $C_{\rm reg}$ corresponds to globally regular configurations, whereas the unshaded region above it is branch-obstructed. The full physical branch $C_{\rm tri}$ remains inside the globally regular region.}
\label{fig4}
\end{figure*}
For the regular geometries considered here, the degeneracy order follows from the sign structure of $f\left(r\right)$. Eq.~(\ref{eq45}) gives $f\left(r\right)>0$ near the center, while asymptotic flatness gives $f\left(r\right)\to1$ at large radius. If the outer event horizon $r_+$ is a simple zero, $f\left(r\right)$ becomes negative immediately inside it. Returning to the static central region requires another sign change at $r_-$, so the inner horizon must be a zero of odd order. The condition $\kappa_-=0$ excludes a simple zero. With a nondegenerate outer horizon, a triple zero is therefore the lowest-order possibility~\cite{CarballoRubio:2022InnerExtremal,DiFilippo:2025InnerExtremal,FrolovZelnikov:2026QTGMassInflation,
DiFilippo:2026QTGMassInflation}.

Accordingly, the triple-degenerate inner horizon considered here satisfies
\begin{align}
\label{eq64}
&f(r_-)=f'(r_-)=f''(r_-)=0,
\quad
f'''(r_-)\neq0,
\nonumber\\
&f(r_+)=0,
\quad
f'(r_+)\neq0,
\quad
0<r_-<r_+.
\end{align}
Near the inner horizon, the metric function has the local behavior
\begin{align}
\label{eq65}
f\left(r\right)=\frac{f'''(r_-)}{3!}(r-r_-)^3
+\mathcal O\left[(r-r_-)^4\right].
\end{align}
A triple zero still allows $f\left(r\right)$ to change sign across the inner horizon, while $f'(r_-)=0$ implies vanishing surface gravity. Hence,
\begin{align}
\label{eq66}
\kappa_-=0,
\quad
\kappa_+=\frac{1}{2}\left|f'(r_+)\right|
\neq0.
\end{align}
This differs from the extremal configurations on $C_{\rm ext}$, where the inner and outer horizons merge and $T_{\rm H}=0$. For a triple-degenerate inner horizon with a simple outer event horizon, only the inner horizon is degenerate. Hence $\kappa_-=0$ while $T_{\rm H}\neq0$.

Since $\rho=br^{2}/Q$ is monotonic for $r>0$, the conditions in Eq.~(\ref{eq64}) can be expressed directly in terms of the horizon curve $A_{\rm h}(\rho,\beta,\xi)$ as
\begin{align}
\label{eq67}
&A=A_{\rm h}(\rho_-,\beta,\xi),
\quad
\left.\frac{\partial A_{\rm h}}{\partial\rho}
\right|_{\rho_-}=0,
\nonumber\\
&\left.\frac{\partial^2 A_{\rm h}}{\partial\rho^2}
\right|_{\rho_-}=0,
\quad
\left.\frac{\partial^3 A_{\rm h}}{\partial\rho^3}\right|_{\rho_-}\neq0,
\end{align}
where $\rho_-=br_-^2/Q$. Unlike the ordinary local minimum of $A_{\rm h}\left(\rho\right)$ associated with $C_{\rm ext}$, a triple-degenerate inner horizon corresponds to a stationary inflection point of the horizon curve. This condition further restricts the combinations of parameters that can support a higher-order degenerate inner horizon. To express the resulting parameter relations, it is convenient to introduce
\begin{align}
\label{eq68}
\chi_- = \frac{\xi\rho_-}{3\beta}=\left(\frac{r_-}{\ell}\right)^2,
\end{align}
which directly measures the inner-horizon radius in units of $\ell$. Using Eq.~(\ref{eq35+}) together with the first- and second-derivative conditions in Eq.~(\ref{eq67}), the triple-degeneracy condition reduces to
\begin{align}
\label{eq69}
\rho_- \chi_-^2&+\left(4{\rho_-} -\sqrt{1+\rho_-^2}\right)\chi_-
\nonumber\\
&+ 3\left(\rho_-+\sqrt{1+\rho_-^2}\right)=0.
\end{align}
This equation gives two formal branches,
\begin{align}
\label{eq70}
\chi_-^{(\pm)}\!=\!\frac{\sqrt{1\!+\!\rho_-^2}\!-\!4\rho_- \!\pm\! \sqrt{1\!+\!5\rho_-^2\!-\!20\rho_-\sqrt{1\!+\!\rho_-^2}}}{2\rho_-}.
\end{align}
\begin{figure*}[t]
\centering
\begin{subfigure}[t]{0.45\textwidth}
\centering
\includegraphics[width=\linewidth]{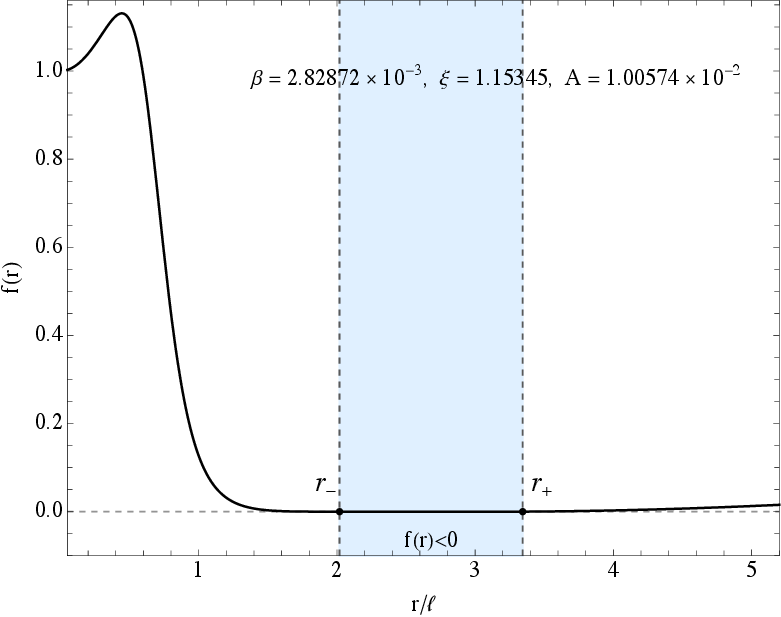}
\caption{}
\label{fig5a}
\end{subfigure}
\hfill
\begin{subfigure}[t]{0.48\textwidth}
\centering
\includegraphics[width=\linewidth]{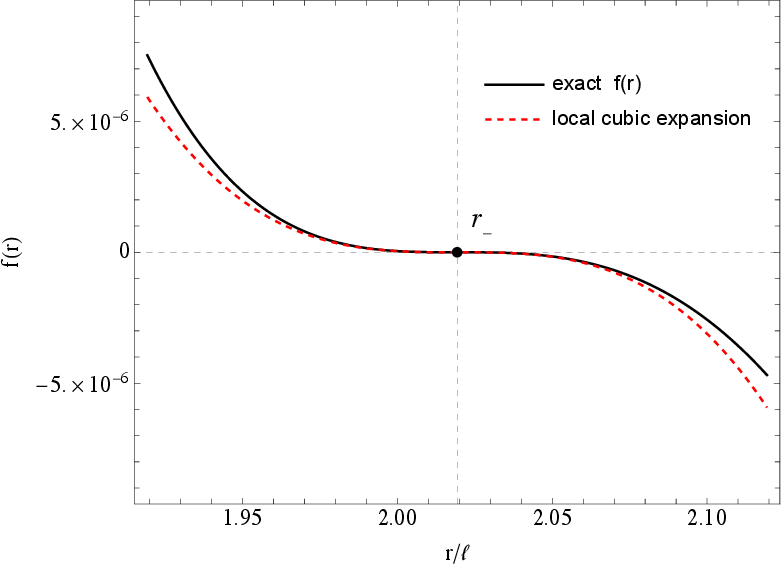}
\caption{}
\label{fig5b}
\end{subfigure}
\caption{Representative RBH with a triple-degenerate inner horizon. (a) Full radial profile of the metric function. The vertical dashed lines mark the triple-degenerate inner horizon $r_-$ and the simple outer event horizon $r_+>r_-$. The shaded interval corresponds to $r_-<r<r_+$, where $f\left(r\right)<0$. (b) Enlarged view of the neighborhood of $r_-$. The black solid curve is the exact metric function, while the red dashed curve is the cubic approximation $f\left(r\right)\simeq f'''(r_-)(r-r_-)^3/3!$.}
\label{fig5}
\end{figure*}
Real solutions require a nonnegative discriminant in Eq.~(\ref{eq70})
and $\chi_->0$. The remaining parameters are then fixed by
\begin{subequations}
\label{eq71}
\begin{align}
\xi_{\rm tri}&=\frac{\chi_-(3+\chi_-)}{\left(\sqrt{1+\rho_-^2}-\rho_-\right)
(1+\chi_-)^2},
\\
\beta_{\rm tri}&=\frac{\xi_{\rm tri}\rho_-}{3\chi_-},
\\
A_{\rm tri}&=\frac{3}{2}\beta_{\rm tri}J_4(\rho_-)+\frac{\rho_-^{3/2}}{1+\chi_-}.
\end{align}
\end{subequations}
Eqs.~(\ref{eq70})-(\ref{eq71}) provide a parametric representation of the triple-degenerate solutions. Since these conditions simultaneously constrain $\beta$, $\xi$, and $A$, a triple-degenerate inner horizon is not a generic property of RBHs, but corresponds to a special subset of parameter space. Evaluating the two formal branches in Eq.~(\ref{eq70}) shows that the physical inner-horizon solutions belong to the $\chi_-^{(-)}$ branch. On this branch, $0<\rho_-<\rho_-^{*}$, with $\rho_-^{*}\simeq0.0421631$, where the upper endpoint is determined by $\partial_{\rho}^{3}A_{\rm h}|_{\rho_-^{*}}=0$. Throughout this interval, the solutions satisfy $A_{\rm ext}(\beta_{\rm tri},\xi_{\rm tri})<A_{\rm tri}<A_{\rm reg}(\beta_{\rm tri})$ and admit a larger simple outer horizon $r_+>r_-$. The resulting physical solutions define the triple-degenerate branch $C_{\rm tri}$. Its parameter-space structure is shown in Fig.~\ref{fig4}. A similar parameter tuning also appears in the construction of inner-extremal RBHs in QTG~\cite{DiFilippo:2025InnerExtremal}.

Fig.~\ref{fig4a} displays the projection of $C_{\rm tri}$ onto the $(\beta,\xi)$ plane. The point with $\rho_-=0.03$ is marked for the representative geometry discussed below. Fig.~\ref{fig4b} shows the position of $C_{\rm tri}$ relative to the global regularity boundary $C_{\rm reg}$. The full physical branch remains below $C_{\rm reg}$, consistently with $A_{\rm tri}<A_{\rm reg}(\beta_{\rm tri})$. Together with $A_{\rm ext}(\beta_{\rm tri},\xi_{\rm tri})<A_{\rm tri}$ established above, this places $C_{\rm tri}$ inside the RBH region rather than on either boundary.

For this representative point, $\rho_-=0.03$, Eq.~(\ref{eq71}) gives $\beta\simeq2.82872\times10^{-3}$,
$\xi\simeq1.15345$, and $A\simeq1.00574\times10^{-2}$.
The corresponding inner and outer horizons are located at $r_-/\ell\simeq2.01932$ and $r_+/\ell\simeq3.34401$, respectively. Their global and near-inner-horizon structures are displayed through the metric function in Fig.~\ref{fig5}.

Fig.~\ref{fig5a} shows the expected sign structure, $f>0$ for $r<r_-$, $f<0$ for $r_-<r<r_+$, and $f>0$ outside $r_+$. The outer root is crossed with nonzero slope and is therefore simple. Near $r_-$, Fig.~\ref{fig5b} shows agreement between the exact metric function and the cubic expansion in Eq.~(\ref{eq65}), providing a numerical check of the triple-zero behavior. The representative solution therefore contains a triple-degenerate inner horizon with $\kappa_-=0$ and a distinct simple outer horizon with $\kappa_+\neq0$.

The vanishing inner-horizon surface gravity may also affect the internal dynamics of this branch. Previous studies of inner-extremal RBHs suggest that the standard exponential blueshift may be modified when $\kappa_-=0$ while the outer horizon remains nondegenerate~\cite{CarballoRubio:2022InnerExtremal,DiFilippo:2025InnerExtremal}. Whether this also suppresses mass inflation requires a separate dynamical analysis.

\subsection{Hawking temperature} 
\label{sec4-4} 
A general thermodynamic formulation has recently been developed for black holes realized as quasi-topological vacuum solutions~\cite{Borissova:2026rbi}. In the present matter-coupled model, we restrict attention to the Hawking temperature associated with the simple outer event horizon. Evaluating Eq.~(\ref{eq55}) on the outer-horizon branch defined by Eq.~(\ref{eq58}) and using Eq.~(\ref{eq57}) gives
\begin{align} 
\label{eq72} 
 T_{\rm H} &= \frac{\sqrt{3\beta}} {4\pi\xi^{3/2}\rho_+^{5/2}\ell} \left[ \rho_+(9\beta+\xi\rho_+) \right.
\nonumber\\
&\left. - (3\beta+\xi\rho_+)^{2} \left(\sqrt{1+\rho_+^{2}}-\rho_+\right) \right], 
\end{align} 
where $\rho_+$ is the larger positive solution of Eq.~(\ref{eq58}).

\begin{figure}[ht]
\centering
\includegraphics[width=1.0\linewidth]{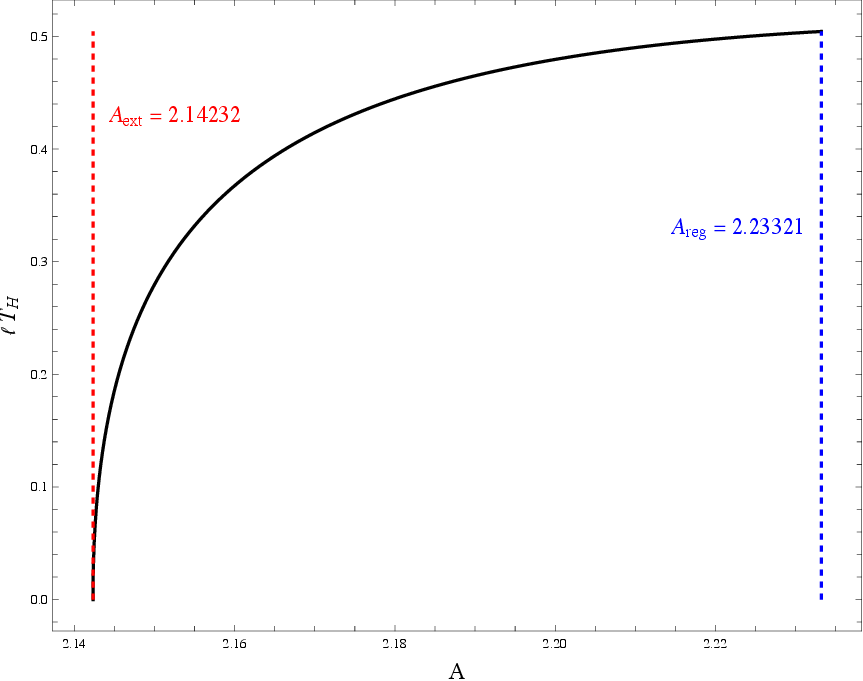}
\caption{Dimensionless Hawking temperature $\ell T_{\rm H}$ along the RBH branch for the representative slice $\beta=\xi=1$.
The temperature vanishes at $A_{\rm  ext}\simeq2.14232$ and remains finite as $A\to A_{\rm reg}^{-}$, where $A_{\rm reg}\simeq2.23321$.}
\label{fig6}
\end{figure}
Fig.~\ref{fig6} shows the Hawking temperature along the representative slice $(\beta,\xi)=(1,1)$. The temperature vanishes at $A=A_{\rm ext}$ and increases monotonically throughout the RBH region $A_{\rm ext}<A<A_{\rm reg}$. It remains finite as $A\to A_{\rm reg}^{-}$, showing that the loss of global regularity at $C_{\rm reg}$ does not coincide with a zero-temperature limit. By contrast, the triple-degenerate branch $C_{\rm tri}$ has a simple outer horizon and therefore $T_{\rm H}\neq0$.

\section{Conclusions and discussion}
\label{sec5}

In this work, we investigated static, spherically symmetric charged solutions in four-dimensional NPQTG coupled to Born-Infeld NED. For the characteristic function $h\left(p\right)=p/(1+\ell^{2}p)$ considered here, the corresponding vacuum branch reaches the critical value at which the characteristic relation becomes singular at a finite radius, leading to a curvature singularity. The exact charged solution shows that, within an appropriate parameter range, the Born-Infeld nonlinearity prevents $H_{\rm BI}\left(r\right)$ from reaching this critical value, allowing the spacetime to extend from an asymptotically flat exterior to a finite-curvature AdS-type core. The resulting regular geometry violates both the timelike and null convergence conditions in a neighborhood of the core, while these conditions are satisfied asymptotically. In contrast to cases in which introducing charge spoils the regularity of a vacuum black hole, Born-Infeld electrodynamics here removes the finite-radius singularity of a gravitational model that is singular in vacuum and supports globally regular charged geometries, including RBHs.

Global regularity is controlled by the condition $A<A_{\rm reg}\left(\beta\right)$. Within this regular sector, the degenerate-horizon boundary $C_{\rm ext}$ separates regular horizonless configurations from RBHs with distinct inner and outer horizons. For the physical parameter range $\beta>0$ and $\xi>0$, $C_{\rm ext}$ lies strictly below the global regularity boundary $C_{\rm reg}$. The RBHs therefore occupy a parameter region of nonzero width rather than a fine-tuned locus. Global regularity and horizon formation are distinct requirements. The former excludes curvature singularities throughout the radial domain, while the latter determines whether a globally regular solution possesses black-hole horizons.

A triple-degenerate inner-horizon branch $C_{\rm tri}$ was also identified within the RBH region. Along this branch, the inner horizon is a triple zero with $\kappa_-=0$, while the outer event horizon remains simple with $\kappa_+\neq0$. Unlike the extremal configurations on $C_{\rm ext}$, the inner and outer horizons remain distinct and the Hawking temperature is nonzero on $C_{\rm tri}$. Previous studies of inner-extremal RBHs suggest that the standard exponential blueshift may be modified when $\kappa_-=0$ while the outer horizon remains nondegenerate. Whether this also suppresses mass inflation in the present higher-curvature model cannot be determined from the static horizon structure alone. Recent numerical studies of perturbed charged black hole interiors have revealed nontrivial critical behavior in the nonlinear development of interior singularities~\cite{Shao:2025fki,Shao:2025apr}, further motivating a dynamical analysis of the present solutions.

The null-geodesic structure of the regular horizonless sector also deserves further study. Regular  horizonless configurations can support stable and unstable light rings~\cite{Eichhorn:2025pgy}, and a systematic analysis of such structures in the present model is left for future work.

\acknowledgments
We thank Johanna Borissova for helpful suggestions and discussions.

\appendix
\section{Explicit expressions for the curvature invariants}
\label{app-0}

For completeness, we collect here the explicit expressions for the
Ricci scalar and the Kretschmann scalar associated with the exact
metric function in Eq.~(\ref{eq24+}). They are given by
\begin{subequations}
\label{eq-app-curvature}
\begin{align}
\label{eq-app0-1}
R\left(r\right)&= \frac{
16\pi G b^{2}\left[\frac{r^{2}}{\Sigma\left(r\right)}
+\frac{\Sigma\left(r\right)}{r^{2}}-2\right]-12H_{\rm BI}\left(r\right)}
{D^{2}\left(r\right)}
\nonumber\\
&+\frac{2\ell^{2}\left\{8\pi G b^{2}\left[
\frac{\Sigma\left(r\right)}{r^{2}}-1\right]
-3H_{\rm BI}\left(r\right)\right\}^{2}}
{D^{3}\left(r\right)}
\nonumber\\
&+\frac{12H_{\rm BI}\left(r\right)}{D\left(r\right)},
\\
K\left(r\right)&=\left[
\frac{2H_{\rm BI}\left(r\right)}{D\left(r\right)}
-\frac{16\pi GQ^2}{r^2\Sigma\left(r\right)D^2\left(r\right)}
\right.
\nonumber\\
&\left.
+\frac{2\ell^2\left\{8\pi G b^2\left[
\dfrac{\Sigma\left(r\right)}{r^2}-1\right]
-3H_{\rm BI}\left(r\right)\right\}^{2}}
{D^3\left(r\right)}
\right]^2
\nonumber\\
&+4\left[
\frac{2H_{\rm BI}\left(r\right)}{D\left(r\right)}
+\frac{8\pi G b^2\left[
\dfrac{\Sigma\left(r\right)}{r^2}-1\right]
-3H_{\rm BI}\left(r\right)}
{D^2\left(r\right)}
\right]^2
\nonumber\\
&+4\left[
\frac{H_{\rm BI}\left(r\right)}{D\left(r\right)}
\right]^2,
\end{align}
\end{subequations}
where $D\left(r\right)=1-\ell^2H_{\rm BI}\left(r\right)$ and $\Sigma\left(r\right)=\sqrt{r^4+{Q^2}/{b^2}}$.

\section{Action-level behavior of the globally regular solutions}
\label{app1}
For completeness, we examine the action-level behavior of the globally regular solutions at the two values of $p$ that require care, the central limit $p\to-1/\ell^{2}$ and the finite-radius crossing $p=0$. The characteristic function is
\begin{align}
\label{eq-app1-1}
H\left(p\right)=6h\left(p\right)=\frac{6p}{1+\ell^{2}p}.
\end{align}
The corresponding coefficient functions are given in Eq.~(\ref{eq7H}). We first consider the central limit. Defining $\delta=1+\ell^{2}p$, their behavior near the non-polynomial boundary $\delta\to0$ is
\begin{subequations}
\label{eq-app1-2}
\begin{align}
H_{2}&=\mathcal O(\delta^{-2}),
\quad
H_{3}=\mathcal O(\delta^{-2}),
\\
H_{4}&=\mathcal O(\delta^{-1})+\mathcal O(\ln\delta),
\quad
H_{4}'=\mathcal O(\delta^{-2}).
\end{align}
\end{subequations}
Thus, the individual coefficient functions of the covariant action become singular as $p\to-1/\ell^{2}$.

For the globally regular solutions obtained in
Section~\ref{sec3-3},
\begin{align}
\label{eq-app1-3}
p\left(r\right)=-\frac{1}{\ell^{2}}
-\frac{r^{3}}{\ell^{4}\Delta}
+\mathcal O(r^{4}),
\quad
\Delta<0,
\end{align}
so that
\begin{align}
\label{eq-app1-4}
\delta = -\frac{r^{3}}{\ell^{2}\Delta}
+\mathcal O(r^{4})=\mathcal O(r^{3}).
\end{align}

The spherically reduced gravitational Lagrangian of the single-function NPQTG theory can be written, up to the overall angular and temporal factors,  as~\cite{Borissova:2026Regular4D}
\begin{align}
\label{eq-app1-5}
L_{\rm grav} & = \frac{{\rm d}}{{\rm d}r}\left[\frac{(1-f)f'}{6}
\int {\rm d}p\frac{H'\left(p\right)}{p^{2}}\right]
\nonumber\\
&+\frac{{\rm d}}{{\rm d}r}
\left[\frac{r^{3}}{3}H\left(p\right)\right],
\end{align}
where the gauge $N=1$ has been used. Near the center, $H\left(p\right)=\mathcal O(\delta^{-1})=\mathcal O(r^{-3})$, and hence $r^{3}H\left(p\right)=\mathcal O(1)$. Furthermore,
\begin{align}
\label{eq-app1-5+}
\int {\rm d}p\frac{H'\left(p\right)}{p^{2}}=\mathcal O(\delta^{-1})+\mathcal O(\ln\delta),
\end{align}
while the regular central metric in Eq.~(\ref{eq45}) gives
\begin{align}
\label{eq-app1-5++}
1-f=\mathcal O(r^{2}), 
\quad
f'=\mathcal O\left(r\right),
\end{align}
which leads to,
\begin{align}
\label{eq-app1-6}
(1-f)f' \int {\rm d}p\,\frac{H'\left(p\right)}{p^{2}}
= \mathcal O(1)+\mathcal O(r^{3}\ln r).
\end{align}
Both quantities within the total derivatives in
Eq.~(\ref{eq-app1-5}) remain finite at the center, and the resulting radial action is locally integrable.

This result is consistent with the finite-curvature central geometry found in Section~\ref{sec3-3},
\begin{align}
\label{eq-app1-6+}
R(0)=-\frac{12}{\ell^{2}},
\quad
K(0)=\frac{24}{\ell^{4}}.
\end{align}
Thus, although the individual non-polynomial coefficient functions are singular in the central limit, the corresponding static,
spherically symmetric solution remains regular and the reduced action has no nonintegrable singularity at $r=0$.

We next consider the finite-radius crossing. As shown in Section~\ref{sec3-2}, the globally regular positive-mass solutions considered here cross $\hat h=0$ at a finite radius. According to Eq.~(\ref{eq32}), this corresponds to $p=0$. Denoting the crossing radius by $r=r_0$, Eq.~(\ref{eq35+}) shows that
the zero is simple, since $\left.
({{\rm d}\hat h}/{{\rm d}\rho})
\right|_{\rho_0}=-{3\beta}J_4'(\rho_0)/{2\rho_0^{3/2}}>0$. Hence $p=\mathcal O(r-r_0)$ near the crossing. The functions $H_2$,
$H_3$, and $H_4$ remain finite there, whereas
\begin{subequations}
\label{eq-app1-7+}
\begin{align}
H_{4}'\left(p\right)
&=\ell^{2}\left[
2\ln|\ell^{2}p|+3+\mathcal O(\ell^{2}p)
\right],
\\
\int {\rm d}p\,\frac{H'\left(p\right)}{p^{2}}
&=-\frac{6}{p}\left[
1+2\ell^{2}p\ln|\ell^{2}p|
+\mathcal O(\ell^{2}p)
\right].
\end{align}   
\end{subequations}
Although $H_{4}'\left(p\right)$ diverges logarithmically, the apparent pole in the first total-derivative term of Eq.~(\ref{eq-app1-5}) cancels. Indeed, using $1-f=r^{2}p$ gives
\begin{align}
\label{eq-app1-8+}
\frac{(1-f)f'}{6}
&\int {\rm d}p\frac{H'\left(p\right)}{p^{2}}
\nonumber\\
&=-r^{2}f'\left[1+2\ell^{2}p\ln|\ell^{2}p| +\mathcal O(\ell^{2}p)\right].
\end{align}
This expression has a finite and continuous limit because $p\ln|p|\to0$. Its radial derivative is at most logarithmically divergent and is therefore locally integrable. The second total-derivative term in Eq.~(\ref{eq-app1-5}) is regular at the crossing because $H\left(p\right)$ is analytic at $p=0$. The reduced bulk equations also remain finite there, since both $H\left(p\right)$ and $H'\left(p\right)$ are regular at $p=0$. Hence, the finite-radius crossing introduces no nonintegrable singularity into the spherically reduced action.

\section{Separation between the regularity and extremal boundaries}
\label{app2}
In Section~\ref{sec4-2}, the regularity boundary $C_{\rm reg}$ and the extremal boundary $C_{\rm ext}$ were introduced to characterize two different properties of the solution. The former determines whether $H_{\rm BI}\left(r\right)$ avoids the critical value
at which the characteristic relation becomes singular, whereas the
latter determines the onset of black hole horizons within the globally
regular solution space. Here we show that these two boundaries remain strictly separated throughout the physical parameter range. This ensures that the RBH region displayed in  Fig.~\ref{fig3a} is not a special feature of the representative choice $\xi=1$.

For fixed $(\beta,\xi)$, the horizon curve is
\begin{align}
\label{eq-app2-1}
A_{\rm h}(\rho,\beta,\xi) = \frac{3}{2}\beta J_4\left(\rho\right)+\frac{3\beta\rho^{3/2}}
{3\beta+\xi\rho}.
\end{align}
The extremal boundary corresponds to the minimum of this curve that controls the transition from a horizonless configuration to a two-horizon black hole. Therefore,
\begin{align}
\label{eq-app2-2}
A_{\rm ext}(\beta,\xi) \leq A_{\rm h}(\rho_{\rm c},\beta,\xi),
\end{align}
where
\begin{align}
\label{eq-app2-3}
\rho_{\rm c} = \frac{\beta}{\sqrt{1+2\beta}}
\end{align}
is the radial position that determines the regularity boundary in
Section~\ref{sec3-2}.

At this point, the difference between the regularity boundary and the horizon curve takes the simple form
\begin{align}
\label{eq-app2-4}
A_{\rm reg}\left(\beta\right) - A_{\rm h}(\rho_{\rm c},\beta,\xi) = \frac{\xi\rho_{\rm c}^{5/2}}
{3\beta+\xi\rho_{\rm c}}.
\end{align}
Since $\rho_{\rm c}>0$ for $\beta>0$, the right hand side is strictly positive for the physical parameter range $\beta>0$ and $\xi>0$.
Consequently,
\begin{align}
\label{eq-app2-5}
A_{\rm ext}(\beta,\xi) < A_{\rm reg}\left(\beta\right),
\quad
\beta>0, 
\quad
\xi>0.
\end{align}
This result shows that the formation of a degenerate horizon always occurs before the solution reaches the boundary at which global
regularity is lost. Hence, for every physical pair $(\beta,\xi)$ there exists a nonzero interval
\begin{align}
\label{eq-app2-6}
A_{\rm ext}(\beta,\xi) < A < A_{\rm reg}\left(\beta\right)
\end{align}
in which the solution is both globally regular and possesses separated inner and outer horizons. The RBH sector is therefore not restricted to a fine-tuned locus in parameter space. Fig.~\ref{fig3a} should be understood as the $\xi=1$  representative slice of this more general result.

%\acknowledgments

\bibliographystyle{apsrev4-1}
\bibliography{template}

%\begin{thebibliography}{}
%\end{thebibliography}
\end{document}